# Achieving High Curie Temperature in (Ga,Mn)As

M Wang, R P Campion, A W Rushforth, K W Edmonds, C T Foxon, and B L Gallagher

School of Physics and Astronomy, University of Nottingham, University Park, Nottingham NG7 2RD, United Kingdom

*We study the effects of growth temperature, Ga:As ratio and post-growth annealing procedure on the Curie temperature, $T_C$, of (Ga,Mn)As layers grown by molecular beam epitaxy. We achieve the highest $T_C$ values for growth temperatures very close to the 2D-3D phase boundary. The increase in $T_C$, due to the removal of interstitial Mn by post growth annealing, is counteracted by a second process which reduces $T_C$ and which is more effective at higher annealing temperatures. Our results show that it is necessary to optimize the growth parameters and post growth annealing procedure to obtain the highest $T_C$.*

75.50.Pp

(Ga,Mn)As is one of the most widely studied dilute magnetic semiconductor systems exhibiting carrier mediated ferromagnetism. For this material to be useful in device applications it will be necessary to increase the Curie temperature ($T_C$) above room temperature. Theory predicts that $T_C$ is proportional to the magnetic moment density which depends upon the density of substitutional Mn ions, $x_S$ [1,2]. This trend has been confirmed experimentally for samples with $x_S<6.8\%$ grown by molecular beam epitaxy (MBE) with $T_C$ reaching 173K[2]. $x_S\approx6.8\%$ is achieved for total Mn concentration $x_{total}\approx9\%$ with the additional $\approx2.2\%$ incorporated as interstitial Mn ($Mn_I$) which can be removed by post-growth annealing[3]. Recent attempts to grow (Ga,Mn)As with larger $x_{total}$ have failed to achieve $T_C$ in excess of the previous record[2] and have produced conflicting results with $T_C$ decreasing[4], saturating[5] or increasing[6] with increasing $x_{total}$. In Ref [5] it was found that $T_C$ saturated at 165K for $x_{total}>10\%$, leading the authors to suggest that the Zener model may not be applicable in the heavily alloyed regime. However, in another study, Olejnik et al[7] reported $T_C=180K$ for $x_{total}=11\%$, obtained by etching and annealing the sample. The range of different results obtained by different groups for $x_{total}>10\%$ indicates that more research is required to understand how growth parameters and post growth annealing procedures affect the achievable $T_C$ when $x_{total}>10\%$. Here we present a detailed study of the epitaxial growth of (Ga,Mn)As layers with $x_{total}\approx12\%$ and $T_C$ up to 185K. We show that the $T_C$ depends sensitively on the growth temperature and the post growth annealing procedures.

25nm (Ga,Mn)As layers were grown by low temperature (~200 °C) MBE on low temperature grown GaAs buffer layers on semi-insulating GaAs (001) substrates, using a Veeco Mod Gen III MBE system. The As flux was provided by a Veeco Mk5 valved cracker, set to produce $As_2$. On a separate test sample, the growth rate was calibrated using RHEED oscillations[8] and the valve setting for As stoichiometry at 580°C was found. The stoichiometric point was found by observing the transition from the As-rich 2x4 to the Ga-rich 4x2 surface reconstruction[9]. The re-evaporation rate of As is higher at 580°C than at ~200 °C, which was found to correspond to ~10% higher As incorporation, allowing a low temperature stoichiometric point to be calculated. Relative adjustments were then made by varying the Beam Equivalent Pressure (BEP).

Excess As or Ga has been shown to lead to the formation of $As_{Ga}$ antisites or 3D growth of GaAs respectively[10]. In agreement with previous studies we find that $T_C$ is highly sensitive to the Ga:As ratio[5,10]. We obtain the highest $T_C$ for a (Ga+Mn):As ratio of 1:1.1 (determined using the method described above), and find that a small deviation from this value leads to a pronounced reduction in the maximum obtainable $T_C$.

Measurement of the substrate temperature $T_g$ was performed using a band edge spectrometer (Bandit from K-Space) under reflection geometry. Typical measurements of $T_g$ during growth are shown in Fig. 1(a) for a series of films with $x_{total}$=12%. The growth temperatures were maintained within ±1°C of the nominal value during the growth of the 25nm (Ga,Mn)As layers (4.5 minutes) by using a low temperature buffer layer and suitable power ramps. Typically $T_g$ remained constant or increased at ≤ 0.5°/min while

2D growth was maintained and then decreased at a similar rate after the 2D-3D phase boundary was crossed, presumably due to a change in surface emissivity.

Each wafer was cut into samples with dimensions 5mm x 4mm which were then annealed at temperatures below the growth temperature (see below). This is an established method[3] for the removal of $Mn_I$ which are detrimental to the ferromagnetism[11]. $T_C$ was determined from measurements of the remnant magnetization in a Quantum Design MPMS SQUID Magnetometer. Each sample was cooled in a field of 0.1T before measuring remnance as a function of increasing temperature in zero applied field.

Figure 1(b) shows $T_C$ for fully annealed samples taken at different distances from the centre of three wafers, with $T_g$ at the start of (Ga,Mn)As growth of 204°C (wafer 1), 202°C (wafer 2), and 200°C (wafer 3). The distance at which the maximum $T_C$ is obtained moves from the edge of the wafer towards the centre as the growth temperature is decreased. Monitoring the RHEED pattern at different points across the wafer indicates that the temperature across the wafer during growth decreases by approximately 2°C from the centre to the edge. These results show that the maximum $T_C$ is very sensitive to the growth temperature, and that the optimum conditions lie very close to the 2D-3D phase boundary which moves closer to the centre of each wafer as the measured substrate temperature is decreased. It is therefore crucial to maintain temperature stability to the accuracy illustrated in Fig. 1(a) in order to obtain the highest possible $T_C$.

Figure 2 shows $T_C$ measured for a series of samples taken from wafer 2 and annealed at temperatures in the range 160°C to 220°C. The anneal time $t_a$ is obtained using the model described in Ref.[3] such that $Dt_a$ is the same as for a sample annealed at 180°C for 48 hours, where $D=D_0 e^{-Q/kT}$ is the diffusivity of $Mn_I$ and $Q=1.4eV$ is the activation energy. We found that these anneal times allowed $T_C$ to reach a saturation value and further annealing did not significantly increase or decrease $T_C$. It is clear that annealing at higher temperature results in a lower $T_C$, suggesting that there is a second process, detrimental to ferromagnetism, which is dependent on temperature. As shown in the inset to Fig. 2, annealing a sample at 220°C after initially annealing at 170°C results in a decrease of $T_C$ to a value similar to samples annealed at 220°C initially. Subsequent annealing at 180°C does not result in an increase of $T_C$ indicating that the second process is not reversible. Further characterization of the samples will be required to elucidate the mechanism responsible for the decrease of $T_C$ at higher annealing temperatures. Possible mechanisms might involve the loss of substitutional Mn to form MnAs precipitates, or to interstitial sites. Other mechanisms might involve the formation of native defects such as $As_{Ga}$ and $Ga_{As}$ antisites which would compensate carriers.

In addition to the wafers described above we have grown wafers with $x_{total}$=10% and 11%. In the absence of moment compensation, $T_C$ should increase roughly linearly with ferromagnetic Mn moment density[2]. Figure 3 shows Curie temperature versus the moment density measured at 2K in zero applied field, for annealed samples showing the highest $T_C$ for a given $x_{total}$ in this and our previous study[2]. For $x_{total} \geq 10\%$ the $T_C$ continues to increase, but with a slightly sub-linear trend. However, we should point out that we consistently achieve higher $T_C$ than previous reports for $x_{total} \geq 10\%$, and it is clear from figure 2 that the annealing procedure has not yet been fully optimized for these samples since $T_C$ shows no sign of saturating as the anneal temperature is reduced.

With increasing $x_{total}$, the growth temperature must be reduced in order to maintain 2D growth, but this increases the probability of forming $As_{Ga}$ antisites and other compensating defects. Hence, precise control over the growth parameters becomes more important for incorporating Mn onto the substitutional sites and for achieving the corresponding increase in $T_C$. We have shown that $T_C$ is extremely sensitive to the growth temperature and (Ga+Mn):As ratio for MBE grown (Ga,Mn)As layers with $x_{total} \geq 10\%$. Additionally, we have found that the post-growth annealing temperature determines the maximum achievable $T_C$ due to a temperature dependent mechanism in addition to the out diffusion of $Mn_I$. These sensitivities may explain why some previous efforts to improve $T_C$ in material with $x_{total} \geq 10\%$ have not been successful.

This work was done in close collaboration with the group of Vit Novak and Tomas Jungwirth (Prague) who have recently also obtained similar high $T_C$ values[12]. We acknowledge funding from EU grant IST-015728 and EPSRC grants GR/S81407/01.

**Figure 1.** (a) Temperature at the centre of three wafers measured by band edge spectrometry as a function of time during the growth of the (Ga,Mn)As layers. The start and end points of the growth of the (Ga,Mn)As layers are indicated by vertical lines. The arrows indicate the approximate point where the 2D-3D RHEED transition occurs. (b) Curie temperature as a function of the distance from the centre of the wafer for wafers 1 (squares) and 2 (circles) annealed at 180$^o$C for 48 hours and wafer 3 (triangles) annealed at 170$^o$C for 116 hours.

**Figure 2** Curie temperature after annealing for 2.6hrs (220$^o$C), 5.2hrs (210$^o$C), 13hrs (200$^o$C), 18hrs (193$^o$C), 48hrs (180$^o$C), 116hrs (170$^o$C) and 280hrs (160$^o$C) for a series of samples taken from wafer 2. Samples were taken from 4 different distances from the centre of the wafer. Inset: TC for a sample taken 7.5mm from the centre annealed at 170$^o$C for 116 hrs then at 220$^o$C for (1) 1 hour, (2) 2 hour and (3) 2 hour intervals, then at 180$^o$C for 48 hours (4).

**Figure 3** Curie temperature versus the spontaneous magnetization measured at 2K for annealed samples showing the highest Tc for a given xtotal. Open squares correspond to the samples with $x_{total} \geq 10\%$ discussed in this paper while closed squares correspond to our previously reported results for $x_{total} <10\%$ [2] which were grown on a different MBE system (Mod Gen II). For comparison, open circles

correspond to samples grown on the Mod Gen III for $x_{total}$ <10% showing that we can achieve similar results on either MBE system.

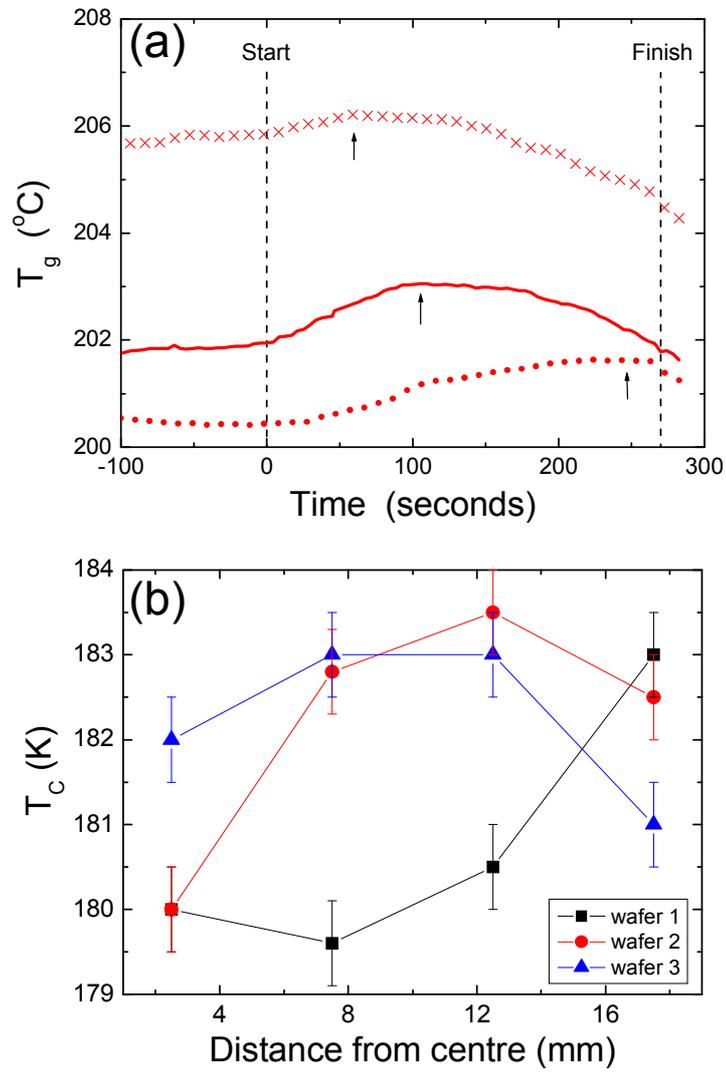

**Figure 1.**

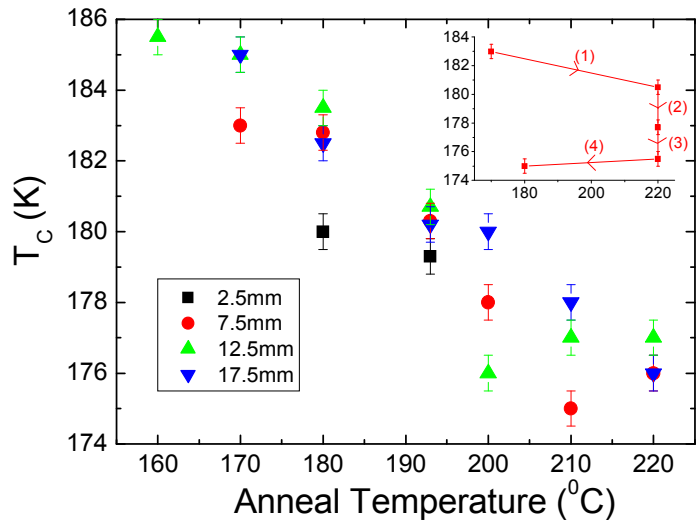

**Figure 2**

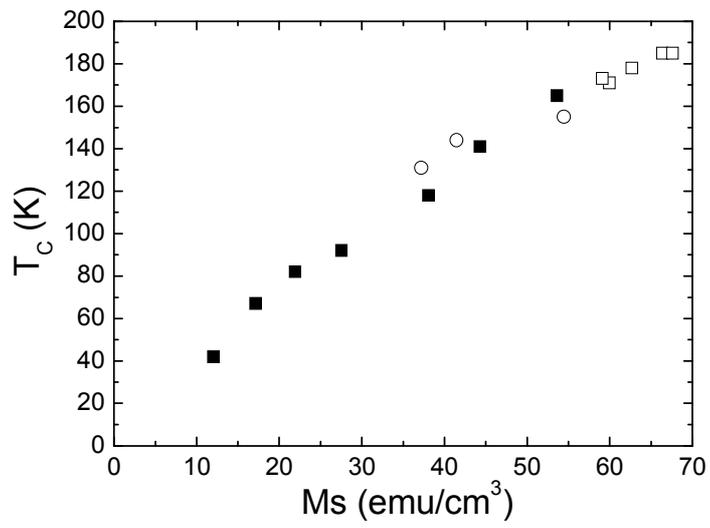

**Figure 3**